\documentclass[useAMS]{mn2e}
\input psfig.sty

\newcommand{\msun}{{\rm M}_{\sun}}
\newcommand{\xte}{{\it RXTE}}
\newcommand{\gro}{{\it CGRO}}

\title{Energy-dependent variability from accretion flows}

\author[A. A. Zdziarski]
{Andrzej A. Zdziarski\thanks{E-mail: aaz@camk.edu.pl}\\
Centrum Astronomiczne im.\ M. Kopernika, Bartycka 18, 00-716 Warszawa, Poland\\
}

\date{Accepted 2005 April 12. Received 2004 December 23}

\pagerange{\pageref{firstpage}--\pageref{lastpage}}
\pubyear{2005}

\begin{document}

\maketitle

\label{firstpage}

\begin{abstract}
We develop a formalism to calculate energy-dependent fractional variability (rms) in accretion flows. We consider rms spectra resulting from radial dependencies of the level of local variability (as expected from propagation of disturbances in accretion flows) assuming the constant shape of the spectrum emitted at a given radius. We consider the cases when the variability of the flow is either coherent or incoherent between different radial zones. As example local emission, we consider blackbody, Wien and thermal Comptonization spectra. In addition to numerical results, we present a number of analytical formulae for the resulting rms. We also find an analytical formula for the disc Wien spectrum, which we find to be a very good approximation to the disc blackbody. We compare our results to the rms spectrum observed in an ultrasoft state of GRS 1915+105.
\end{abstract}

\section{INTRODUCTION}
\label{intro}

A very common tool in studying X-ray sources is spectral fitting, e.g., with models included in the {\sc xspec} packet (Arnaud 1996). In this way, a spectrum averaged over some period can be decomposed into components due to various radiative processes. However, spectral fitting is not unique in general, and the same data can often be well fitted with more than one spectral model. 

Another way of studying X-ray sources is via their power spectra. However, power spectra are usually integrated over a large range of photon energies, which makes it difficult to make inferences regarding the underlying radiative processes. Also, progress towards physical understanding of power spectra of, e.g., X-ray binaries, has been rather modest. The best present physical description of their power spectra appears to be in terms of Lorentzians (e.g., Nowak 2000; Belloni, Psaltis \& van der Klis 2002), Fourier transforms of damped harmonic oscillators. However, this decomposition is rather orthogonal to the decomposition of energy spectra into components due to different radiative processes. 

Yet another observable is the energy-dependent variance, $\sigma_E^2$, or the rms fractional variability, $r_E\equiv \sigma_E/F_E$, where $E$ is the photon energy and $F_E$ is the time-averaged flux as a function of $E$. For a given time bin of the underlying lightcurve and the length of the observation, the variance is an integral of the power spectrum over the frequency range spanning those two time intervals. Alternatively, one can compute the variance corresponding to a narrower range of time scales (frequencies) or to a component in a fit to a power spectrum, e.g., the variance corresponding to a QPO. Given $r_E$ and $F_E$, we can compute the spectrum of the corresponding variable component,
\begin{equation}
F^{\rm var}_E=\sigma_E= r_E F_E.
\label{fvar}
\end{equation}
If $r_E$ is calculated for a range of Fourier frequency, $F^{\rm var}_E$ represents a frequency-resolved spectrum (Revnivtsev, Gilfanov \& Churazov 1999). We note, however, that $F^{\rm var}_E$ may not always be interpreted as a spectrum from a physical process (e.g., for the case of spectral pivoting, see below), while $r_E$ can always be interpreted physically. 

Spectra of rms of astrophysical X-ray sources have often been computed (e.g., M\'endez et al.\ 1997; Lin et al.\ 2000; Edelson et al.\ 2002; Wardzi\'nski et al.\ 2002; Zdziarski et al.\ 2002a, b, 2004, 2005; Vaughan et al.\ 2003; Gilfanov, Revnivtsev \& Molkov 2003; Frontera et al.\ 2003; Vaughan \& Fabian 2004; Fabian et al.\ 2004; Gallo et al.\ 2004; Rodriguez et al.\ 2004). However, interpretation or theoretical models of observed rms spectra have so far been relatively rare. E.g., M\'endez et al.\ (1997) and Gilfanov et al.\ (2003) have fitted $F^{\rm var}_E$ of X-ray binaries by blackbody and Wien spectra, respectively. Then, Vaughan \& Fabian (2004), Fabian et al.\ (2004) and Zdziarski et al.\ (2005) have modelled observed rms spectra due to superposition of different spectral component, each characterized by a different variance (see Section \ref{comp}). Gallo et al.\ (2004) have shown that the rms spectrum of the Seyfert 1H 0707--495 is well reproduced by the ratio of the spectra of its high and low flux levels multiplied by a constant. This result can be simply explained as that ratio being proportional to the range of fluxes at which the source varies at a given energy, which should indeed be proportional to its rms. 

As found by Zdziarski et al.\ (2002b), the dominant cause of the variability in the hard state of Cyg X-1, and probably some other black hole binaries, on certain time scales is the varying flux in the soft seed photons irradiating the Comptonizing plasma and much less varying the power supplied to the hot plasma. This results in a spectral pivoting with the pivot point in the middle of the spectrum, and the rms below the pivot decreasing with energy. Such rms spectra decreasing with energy in the X-ray range in the hard state have been found, e.g., by Wardzi\'nski et al.\ (2002) and Frontera et al.\ (2003). Then, Zdziarski et al.\ (2003) presented a formalism to calculate the rms of a pivoting spectrum under various assumptions regarding the distribution of the spectral indices and the possible spread of the pivot energy. Those authors also show this gives a very good description of the long-term rms of Cyg X-1 in the hard state. 

An opposite variability pattern has been found in the soft state of Cyg X-1. Churazov, Gilfanov \& Revnivtsev (2001) have found the variability pattern in the \xte/PCA data consists of a constant disc blackbody emission and a variable high-energy tail. Zdziarski et al.\ (2002b) have found a very similar pattern in the \gro/BATSE and \xte/ASM data, and have interpreted it as resulting from a hot plasma with a variable power supply irradiated by an approximately constant flux of soft photons. Consistent with this picture, they found the rms spectrum of those data increasing with energy in the X-ray range and saturating at an approximately constant level around $\sim$100 keV. 

Another considered cause of energy-dependent rms is time evolution of flares. \.Zycki (2004) have calculated the rms spectrum using the flare model of Poutanen \& Fabian (1999) and showed it results in the rms increasing with energy. 

Here, we further consider the potential of rms spectra for physical interpretation of X-ray emission of cosmic sources and identifying the underlying radiative processes. We develop a formalism for calculating rms spectra from an accretion flow in which the local level of variability is a function of radius, as, e.g., in the propagation models of Lyubarskii (1997), Kotov, Churazov \& Gilfanov (2001), \.Zycki (2003). In the presented method, we assume that the shape of emission from a given radius is constant, and thus the changes of the rms with energy are due to superposition of emission from different radii. We apply this formalism to the cases of blackbody and e-folded power-law local spectra. We compare the model rms spectra to one observed from the black-hole binary GRS 1915+105.

\section{Variability from different spectral components}
\label{comp}

Let us consider emission spectrum averaged over a certain period of time consisting of a number of components, $F= \sum_i F_{i}$, where $F_{i}$ is the averege flux in the $i$-th spectral component. The variability of each component is given by its standard deviation, $\sigma_i$. For notational simplicity, we omit hereafter the average sign for $F$ and the energy subscript for $\sigma$ and $F$, i.e., $F=\langle F_E\rangle$, $\sigma=\sigma_E$. 

The total variance, $\sigma^2$, has then contributions from all of the components,
\begin{equation}
\sigma^2=\sum_i\ \sigma_i^2+ \sum_{i\neq j} \sigma_{ij},
\end{equation}
where the covariance is $\sigma_{ij}=\sigma_i \sigma_j$ in the case of complete coherence (same phases), and $=0$ for the complete incoherence. Hereafter, we consider only these two limiting cases, in which 
\begin{equation}
\sigma_{\rm coh}=\sum_i \sigma_i, \qquad  \sigma_{\rm inc}^2=\sum_i \sigma_i^2, 
\end{equation}
respectively. Note that $\sigma_{\rm coh}^2 \geq \sigma_{\rm inc}^2$. 

The relative level of variability in each component is $r_i\equiv \sigma_i/F_i$. Then, the total rms is
\begin{equation}
r_{\rm coh}=\sum_i r_i {F_i\over F},\quad\quad
r_{\rm inc}^2=\sum_i r_i^2 {F_i^2\over F^2},
\label{sums}
\end{equation}
in the fully coherent and incoherent case, respectively. Note that even when the variability of various components is coherent and each $r_i$ is independent of energy (but different for different component), the total $r_{\rm coh}$ will change with energy due to varying relative contributions of spectral components. 

\section{Variability from an accretion flow}
\label{disc}

We then consider an accretion disc, with the flux per unit area, $F(R)$, and the fractional variability, $r(R)$, being functions of the radius from $R_{\rm in}$ to $R_{\rm out}$. The flux from the disc per unit angle and energy is 
\begin{equation}
F=2 \upi \int_{R_{\rm in}}^{R_{\rm out}} \!{\rm d}\!R\, R F(R).
\label{flux}
\end{equation}
If the variability is coherent over that range of $R$, 
\begin{equation}
\sigma_{\rm coh} =2 \upi \int_{R_{\rm in}}^{R_{\rm out}} \!{\rm d}\!R\, R r(R) F(R).
\label{coh}
\end{equation}

If the variability is completely incoherent over infinitesimally small ranges of the radius, $\sigma_{\rm inc}^2$ becomes also infinitesimally small due to the effect of averaging over a very large number of small contributions. On the other hand, the variance will be finite when the variability is coherent within some finite regions varying approximately independently of each other. For example, the variability can be coherent within a given coherence length, $\Delta R_i$. Then 
\begin{equation}
\sigma_{\rm inc}^2=4 \upi^2 \sum_i \left[ \int_{\Delta R_i} \!{\rm d}\!R\, R r(R) F(R)\right]^2,
\label{int_sum_lin}
\end{equation}
and hereafter the summation covers the range from $R_{\rm in}$ to $R_{\rm out}$. If $\Delta R_i$ is constant ($=\Delta R$) and small compared to the length over which $F(R)$ varies, we have
\begin{equation}
\sigma_{\rm inc}^2\approx 4 \upi^2 (\Delta R)^2 \sum_{i=0} \left[ R_i r(R_i) F(R_i)\right]^2,
\label{sum_lin}
\end{equation}
where $R_i=R_{\rm in}+(i+1/2)\Delta R$. Note that although equation (\ref{int_sum_lin}) is formally more accurate than the sum of equation (\ref{sum_lin}), it may not be more accurate in reality given the approximate character of the assumption of the sharp boundaries of the independent subregions with constant rms each. We also note that the relative rms of equation (\ref{sum_lin}) should be defined with respect to the flux calculated using the same assumptions, i.e., 
\begin{equation}
F\approx 2 \upi \Delta R \sum_{i=0} R_i F(R_i).
\label{flux_lin}
\end{equation}
The sum (\ref{sum_lin}) can in turn be approximated by
\begin{equation}
\sigma_{\rm inc}^2\approx  4 \upi^2 \Delta R \int_{R_{\rm in}}^{R_{\rm out}} \!{\rm d}\!R\, [R r(R) F(R)]^2.
\label{int_lin}
\end{equation}
Within this approximation, $\sigma_{\rm inc}^2$ scales linearly with $\Delta R$. Note that this approximation fails when only few terms contribute to the sum of equation (\ref{sum_lin}). 

We then consider the case in which the variability is coherent over a constant logarithmic range of radius, $\Delta\!\ln R$. This is likely the case for a self-similar flow, which properties are determined by the local size scale. Then equations (\ref{int_sum_lin}), (\ref{sum_lin}) and (\ref{int_lin}) become
\begin{eqnarray}
\lefteqn{ \sigma_{\rm inc}^2=4 \upi^2 \sum_i \left[ \int_{\Delta\!\ln R_i} \!{\rm d}\!\ln R\, R^2 r(R) F(R)\right]^2, \label{int_sum_log}}  \\ 
\lefteqn{ \quad\, \approx 4 \upi^2 (\Delta\!\ln R)^2 \sum_{i=0} \left[ R_i^2 r(R_i) F(R_i)\right]^2, \label{sum_log}}   \\ 
\lefteqn{ \quad\, \approx 4 \upi^2 \Delta\!\ln R \int_{\ln R_{\rm in}}^{\ln R_{\rm out}} \!{\rm d}\!\ln R\, [R^2 r(R) F(R)]^2, \label{int_log}} 
\end{eqnarray}
where $R_i=R_{\rm in}\exp[(i+1/2)\Delta\!\ln R]$. In the approximation of equation (\ref{int_log}), $\sigma^2$ scales linearly with $\Delta\!\ln R$, which approximation fails, however, when only a few regions dominate the variability. Equation (\ref{flux_lin}) is now replaced by
\begin{equation}
F\approx 2 \upi \Delta\!\ln R \sum_{i=0} R_i^2 F(R_i).
\label{flux_log}
\end{equation}
For comparison, we also calculate the coherent variability using the same approximations as for equations (\ref{sum_log}) and (\ref{flux_log}), which yields
\begin{equation}
\sigma_{\rm coh}\approx 2 \upi \Delta\!\ln R \sum_{i=0} R_i^2 r(R_i) F(R_i).
\label{coh_log}
\end{equation}

\section{Radial dependencies of spectra and the variability amplitude}
\label{radial}

Now we consider some examples of the radiated spectra and the radial dependencies of both the spectra and the local rms, $r(R)$. For an emitted spectrum, we assume its local power is equal to the local dissipation rate, i.e., neglect advection. We consider two dimensionless forms, $D(R)$, of the dissipation rate,
\begin{eqnarray}
\lefteqn{ D(R)=(R/R_{\rm in})^{-3}, \label{d_no_b}}  \\ 
\lefteqn{ D(R)= {7^7 \left[1-(R_{\rm in}/R)^{1/2}\right] \over 6^6 (R/R_{\rm in})^3, } \label{d_b}}
\end{eqnarray}
where $R_{\rm in}$ is the inner disc radius. The first dependence corresponds 
to a Newtonian disc torqued at the inner boundary (as assumed, in particular, in the commonly-used disc blackbody model, {\tt diskbb} in {\sc xspec}, Mitsuda et al.\ 1984).  It corresponds, e.g., to either $R_{\rm in}\gg$ the minimum stable orbit (which is between $6GM/c^2$ and $\sim 1GM/c^2$ from nonrotating to the maximally rotating black hole) or the magnetic stresses important near the minumum stable orbit (e.g., Merloni \& Fabian 2003 and references therein). The second dependence corresponds to a nonrotating black hole with the standard zero-torque inner boundary and $R_{\rm in}=6GM/c^2$ (Shakura \& Sunyaev 1973). Both expressions are normalized to the maximum of $D=1$, reached at $R/R_{\rm in}=1$ and $(7/6)^2$, respectively. 

We then consider two simple local spectral forms, a blackbody and a power law with both low and high-energy cutoffs, which represents a rough approximation to spectra from thermal Comptonization. They correspond, e.g., to the radiative processes dominant in the soft and hard spectral states of black hole binaries, respectively (e.g., Zdziarski \& Gierli\'nski 2004). 

The blackbody spectrum is
\begin{equation}
F(R)={A E^3 \over \exp[E/kT(R)]-1},
\label{fbb}
\end{equation}
where $A$ is the normalization constant, $k$ is the Boltzmann constant, and 
\begin{equation}
T(R)=T_{\rm max} \left[ D(R)\right]^{1/4}
\label{temp}
\end{equation}
is the local temperature expressed via its maximum, $T_{\rm max}$. We also consider the Wien limit of the blackbody spectrum, see Appendix A.

Then we consider a power law with an exponential cutoff, which roughly approximates emission due to unsaturated Comptonization from a hot plasma. As in the blackbody case, we normalize the spectrum to the local dissipation rate,
\begin{equation}
F(R)= {A D(R) E^{-\alpha(R)} \exp(E/E_{\rm c})\over 
E_{\rm c}^{1-\alpha(R)} \Gamma \left[ 1-\alpha(R),  E_0/E_{\rm c}\right]}, \quad E\geq E_0,
\label{fpl}
\end{equation}
and $F=0$ for $E<E_0$, where $\Gamma $ is the incomplete Gamma function [$\Gamma(a,x)\rightarrow \Gamma(a),\, x\ll 1$, $\Gamma(a,x)\rightarrow x^{a-1} \exp(-x),\, x\gg 1$] $\alpha(R)$ is the local photon index, $E_{\rm c}$ is the e-folding energy, the dimensional constant, $A$, depends on the central mass and the accretion rate, and $E_0$ is the minimum energy in the spectrum, roughly corresponding to the average energy of the seed photons undergoing Comptonization. 

We then consider a model with a hot inner flow surrounded by a cold disc, providing the seed photons (e.g., Zdziarski et al.\ 2002). Then the flux in the seed photons decreases with the decreasing radius and the locally-emitted spectrum hardens. This is confirmed by Fourier-resolved spectroscopy, showing, e.g., the spectrum of Cyg X-1 in the hard state hardens with the increasing Fourier frequency (Revnivtsev et al.\ 1999). Since the Fourier-resolved spectrum has the slope depending on the logarithm of the Fourier frequency, we assume here $\alpha(R)$ to be linear in the logarithm of radius,
\begin{equation}
\alpha(R)=\alpha_{\rm in}+{\ln (R/R_{\rm in})\over \ln (R_{\rm out}/R_{\rm in})}(\alpha_{\rm out}- \alpha_{\rm in}),
\label{gamma}
\end{equation}
where the spectral index changes from $\alpha_{\rm out}$ at the outer radius of the hot flow to $\alpha_{\rm in}$ at the inner radius. Note that the rms($E$) dependence for the case of constant $\alpha$ would be constant as well. 

\begin{figure}
\centerline{\psfig{file=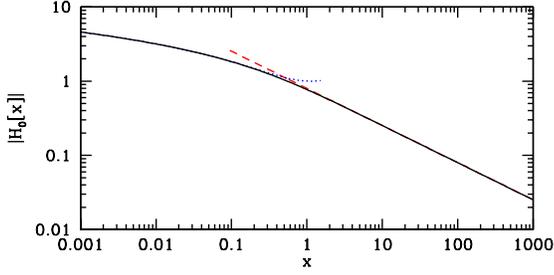,width=7.3cm}} 
\caption{The function $\vert H_0(x)\vert$. The dotted and dashed curves give the approximations for low and high, respectively, values of $x$, given in Section \ref{radial}.
\label{hankel_plot} }
\end{figure}

We then need to specify the fractional variability as a function of the radius. If we consider a perturbation with a given frequency, $f$, propagating with a constant speed, $c_{\rm s}$, towards a sink at the center (or from a central source outside), the wave amplitude will increase with decreasing radius as (see Vaughan \& Nowak 1997; Nowak et al.\ 1999; Misra 2000),
\begin{equation}
r(R)=r_{\rm max} {\vert H_0 (2\upi R/\lambda )\vert \over \vert H_0 ( 2\upi R_{\rm in}/ \lambda)\vert},
\label{hankel}
\end{equation}
where $\lambda=c_{\rm s}/f$ is the wavelength, $H_0$ is the Hankel function of order zero, which modulus is given by
\begin{equation}
\vert H_0(x)\vert^2 = J_0(x)^2+Y_0(x)^2,
\end{equation}
and $J_0$ and $Y_0$ are the Bessel functions of order zero of the first and second kind, respectively. In writing equation (\ref{hankel}), we have assumed that the local rms level, $r(R)$, is proportional to the wave amplitude, and normalized it to its maximum of $r_{\rm max}$ at the inner disk radius. The function $\vert H_0(x)\vert$ is plotted in Fig.\ \ref{hankel_plot}. Its argument is, e.g., unity for $f=3$ Hz, $c_{\rm s}=0.01 c$, and $R\simeq 30GM/c^2$ for $M=10\msun$. At $x\rightarrow 0$, it is logarithmically divergent, $\vert H_0(x)\vert \simeq \{1+(2/\upi)^2[\gamma +\ln (x/2)]^2\}^{1/2}$, and for $x\ga 1$, $\vert H_0(x)\vert \simeq (\upi x/2)^{-1/2}$, see Fig.\ \ref{hankel_plot}. Here $\gamma \simeq 0.577216$ is Euler's constant. 

The above dependence of the wave amplitude as $R^{-1/2}$ can be understood as compression of the energy density of a constant-speed cylindrical wave with decreasing radius, $\propto R^{-1}$. This increase becomes smaller when the wavelength becomes comparable with the radius, see Fig.\ \ref{hankel_plot}. We note that the sound speed in actual accretion discs generally increases with decreasing $R$, which would flatten the dependence of equation (\ref{hankel}). A wave travelling through a geometrically-thin disc has $c_{\rm s}\propto R^{-9/20}$ and $R^{-3/16}$ in the gas-pressure and radiation-pressure dominated cases, respectively (e.g., Svensson \& Zdziarski 1994). This would lead to $r\propto R^{-13/32}$ and $R^{-11/40}$, respectively. The dependence for the gas-pressure dominated case is modified close to the inner radius for the case of the zero-torque boundary, leading to $c_{\rm s}\rightarrow 0$ at $R\rightarrow R_{\rm in}$, which would strongly amplify the wave. Furthermore, wave reflection from the inner boundary is neglected in the dependence of equation (\ref{hankel}). A treatment of those effects is beyond the scope of this paper.

The rms radial profile given by equation (\ref{hankel}) has been obtained for a single frequency. Thus, that rms profile may apply for QPOs. However, the total rms dependence in some states follows that of the low-frequency QPO (Zdziarski et al.\ 2005), and equation (\ref{hankel}) is likely to be applicable more generally. We also note that perturbations propagating via waves will maintain coherence between emission at different radii (Vaughan \& Nowak 1997), and thus this radial rms profile applies to the coherent variability only.

On the other hand, the local fractional variability may be due to superposition of fluctuations (correlated or not) occuring at a number of radii (Lyubarskii 1997). In that case, it will be likely of a power-law form. We assume it here to depend as a power law of either the radius or the dissipation rate,
\begin{eqnarray}
\lefteqn{
r(R)=r_{\rm max} \left( R/R_{\rm in} \right)^{-\beta}, \label{r_pl}}  \\ 
\lefteqn{r(R)=r_{\rm max} \left[ D(R)\right]^{\beta/3}, \label{r_d}}
\end{eqnarray}
 where $\beta$ is the index of the dependence. Note that equation (\ref{r_d}) with $D(R)$ of equation (\ref{d_no_b}) is identical to equation (\ref{r_pl}). The profile of equation (\ref{hankel}) at $2\upi R/\lambda\ga 1$ corresponds to $\beta=1/2$ of equation (\ref{r_pl}). Hereafter, we will consider only the values of $\beta>0$.

\section{Analytical solutions}
\label{analytical}

In the case of blackbody and Wien (see Appendix A) spectra with the torqued dissipation profile, equation (\ref{d_no_b}), and $R_{\rm out}=\infty$ we can obtain a number of analytical solutions for the coherent $\sigma$, equation (\ref{coh}). First, 
\begin{equation}
\sigma_{\rm coh} ={8 \upi\over 3} A E^3 r_{\rm max} R_{\rm in}^2   
{\epsilon - \ln\left(\exp \epsilon -1\right)\over \epsilon },
\label{coh4_5}
\end{equation}
in the case of blackbody emission and $\beta=5/4$. Here $\epsilon \equiv E/kT_{\rm max}$. For blackbody emission at a general value of $\beta$, we find the limiting forms,
\begin{equation}
{\sigma_{\rm coh}\over F}\simeq  r_{\rm max}\cases{ \epsilon^{4\beta/3}
{\zeta \left(8-4\beta\over 3\right) \Gamma  \left(8-4\beta\over 3\right) \over  \zeta (8/ 3)\Gamma (8/3) }, &$\epsilon\ll 1,\, \beta<{5\over 4}$;\cr
{\epsilon^{5/3}\ln(1/\epsilon)\over \zeta (8/ 3)\Gamma (8/3) }, &$\epsilon\ll 1,\, \beta={5\over 4}$;\cr
{3\epsilon^{5/3}\over (4\beta-5) \zeta (8/ 3)\Gamma (8/3) }, &$\epsilon\ll 1,\, \beta>{5\over 4}$;\cr
1, &$\epsilon\gg 1$,\cr}
\label{bb_coh}
\end{equation}
where $\zeta$ is the Riemann zeta function and $\Gamma $ is the Gamma function. Note the power law index of 5/3 at low energies appears independently of the value of $\beta$ as long at it is $>5/4$, whereas it is $4\beta/3$ for $\beta<5/4$.

Then, the corresponding results for a Wien disc are,
\begin{equation}
\sigma_{\rm coh}={8 \upi\over 3} A  r_{\rm max} R_{\rm in}^2 E^3 
\epsilon^{(4\beta-8)/3} \Gamma \left({8-4\beta\over 3}, \epsilon\right),
\label{coh_wien}
\end{equation}
and
\begin{eqnarray}
\lefteqn{\label{ratio_wien}
 {\sigma_{\rm coh}\over F}=r_{\rm max} \epsilon^{4\beta/3} {\Gamma \left({8-4\beta\over 3}, \epsilon\right)\over \Gamma \left({8\over 3}, \epsilon\right)} }\\
\lefteqn{
\label{ratio_wien_limits}
\quad \simeq r_{\rm max} \cases{ \epsilon^{4\beta/3} {\Gamma \left({8-4\beta\over 3}\right)\over \Gamma(8/3)}, &$\epsilon\ll 1$;\cr
1, &$\epsilon\gg 1$.\cr} } 
\end{eqnarray}
Now, the low-energy power law is generally $\epsilon^{4\beta/3}$, in contrast to the blackbody case, where it appears only for $\beta<5/4$. 

In analogous way, we may calculate the dependences for the incoherent $\sigma$ using the integral approximations of either equation (\ref{int_lin}) or (\ref{int_log}). For example, the Wien emission at a logarithmic coherence length yields,
\begin{equation}
 {\sigma_{\rm inc}\over F}={3^{1/2}\over 2^{11/3}} r_{\rm max}(\Delta\! \ln R)^{1/2}  (2\epsilon)^{4\beta/3} {\Gamma^{1/2} \left({16-8\beta\over 3}, 2\epsilon\right)\over \Gamma \left({8\over 3}, \epsilon\right)}.
\label{inc_wien}
\end{equation}
The $\epsilon\ll 1$ limit is obtained trivially by replacing the incomplete Gamma function by the complete one, which yields the same power law index as the corresponding coherent formula (\ref{ratio_wien_limits}). The expression is not valid for $\epsilon\gg 1$, where the actual $\sigma_{\rm inc}/F\rightarrow r_{\rm max}$. This limit also applies to the blackbody emission, which case has then the $\epsilon\ll 1$ limit of
\begin{eqnarray}
\lefteqn{
 {\sigma_{\rm inc}\over F}\simeq r_{\rm max}(\Delta\! \ln R)^{1\over 2} \nonumber}\\
\lefteqn{
\times\cases{
{3^{1\over 2} \epsilon^{4\beta\over 3} \Gamma^{1\over 2} \left(16-8\beta\over 3\right) \left[\zeta\left(13-8\beta\over 3\right)-\zeta\left(16-8\beta\over 3\right)\right]^{1\over 2}
\over 2 \Gamma(8/3) \zeta(8/3)}, &$\beta<{5\over 4}$,\cr 
{3   \epsilon^{5\over 3} \over 2^{3\over 2} (4\beta-5)^{1\over 2} \Gamma(8/3) \zeta(8/3)}, &$\beta>{5\over 4}$.\cr}}
\label{inc_bb}
\end{eqnarray}
It again has the same power law indices as the corresponding coherent cases, equation (\ref{bb_coh}). Finally, $\beta=5/4$ yields,
\begin{eqnarray}\lefteqn{
\sigma_{\rm inc} ={4 \upi\over \sqrt{3}} A E^3 r_{\rm max} R_{\rm in}^2 (\Delta\! \ln R)^{1\over 2} \nonumber}\\
\lefteqn{\qquad \times
\left[  {\ln(\exp \epsilon -1)\over \epsilon } + {1\over \epsilon(\exp\epsilon-1)}-1\right]^{1/2},}
\end{eqnarray}
valid at any $\epsilon$ except $\epsilon \gg 1$. At $\epsilon\ll 1$, the factor in the second line approaches $\epsilon^{-1}$.

\section{Validity of approximations}
\label{validity}

\begin{figure}
\centerline{\psfig{file=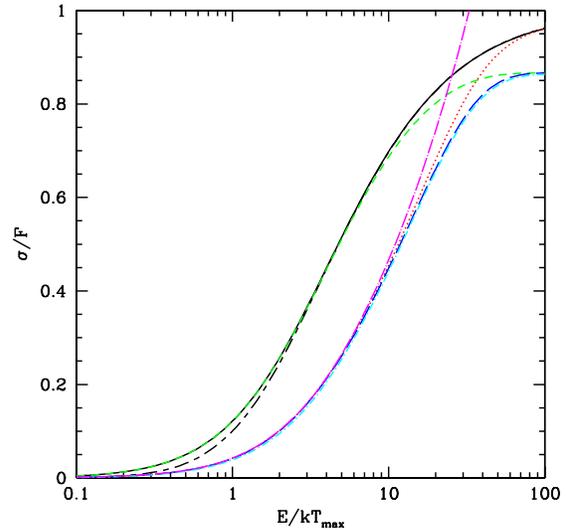,width=7.3cm}} 
\caption{Comparison of various approximations to rms profiles for blackbody emission of a disc with the torqued inner boundary and a power law radial rms profile with $\beta=3$. The solid curve, short/long dashes and short dashes show the case with full coherence, its Wien approximation, and approximation by 
a discrete sum with $\Delta\!\ln R=0.1$, respectively. The dots correspond to the incoherent sum of integrals and $\Delta\!\ln R=0.1$, and the long dashes, its approximation by the discrete sum. The dots/long dashes give in turn the approximation of the latter rms by an integral. The dots/short dashes correspond to the discrete sum with the linear coherence length with $\Delta (R/R_{\rm in})=0.1$. Here and in Figs.\ \ref{rms_beta}--\ref{rms_delta}, the dissipation profile with the torqued inner boundary is used. 
\label{rms_b3} }
\end{figure}

Fig.\ \ref{rms_b3} compares the accuracy of various approximations of Section \ref{disc}, for blackbody emission with the local rms profile of equation (\ref{r_pl}) with $\beta=3$ and the temperature profile of equation (\ref{d_no_b}), i.e., with the torqued inner boundary. Hereafter, $r_{\rm max}=1$, but the rms profiles below scale linearly with it. Also, in order to reduce the number of free parameters, we hereafter assume $R_{\rm out}=\infty$ for blackbody and Wien emission, which assumption has a negligible effect on the shown profiles as long as $R_{\rm out}\gg 10^2 R_{\rm in}$.

The solid curve and the short dashes correspond to the case with full coherence, equations (\ref{coh}) and (\ref{flux}), and its discrete sum approximation by equations (\ref{coh_log}) and (\ref{flux_log}) for $\Delta\!\ln R=0.1$, respectively. We see that the approximation begins to yield $\sigma_{\rm coh}/F$ lower than the integral for $E/kT_{\rm max}\ga 10$. This is due to the discrete sampling of $r$, which results in the maximum sampled $r$ being $<1$ (in the middle of the innermost radial interval). The short/long dashes show the coherent rms in the case of local Wien emission, equation (\ref{ratio_wien}).

The dots correspond to the incoherent variability given by the sum of integrals, equation (\ref{int_sum_log}), at $\Delta\!\ln R=0.1$, and the long dashes, to its approximation by the discrete sums, equations (\ref{sum_log}), (\ref{flux_log}). We see here the analogous effect to the one pointed above, with the discrete approximation sampling only $r<1$, resulting in the rms lower than that due to the sum of integrals for $E/kT_{\rm max}\ga 10$. Then, the dots/long dashes show the rms approximation by equation (\ref{int_log}) with respect to the flux of equation (\ref{flux}). We see it starts to diverge for $E/kT_{\rm max}\ga 10$, which is due to using the integral to approximate the discrete sum (\ref{sum_log}). Still, at $E/kT_{\rm max}\la 10$ (for $\Delta\!\ln R=0.1$), all the methods of calculating the rms yield almost identical results in either the coherent or the inchorent case. Finally, the dots/short dashes correspond to the case with the linear coherence length with $\Delta (R/R_{\rm in})=0.1$ and equations (\ref{sum_lin}--\ref{flux_lin}), which yields almost identical values of the rms as the corresponding logarithmic coherence length. 

\section{Example rms energy profiles}
\label{examples}

Hereafter, we use the fully accurate formulae, i.e., with integrals and their sums. Equations (\ref{coh}) and (\ref{flux}) are used in the coherent case, and equations (\ref{int_sum_log}) and (\ref{flux}) in the incoherent case. Subsequent figures are shown in the logarithmic scale of $\sigma/F$, in which case the shape of a dependence is invariant with respect to changes of the normalization, $r_{\rm max}$. Also, it allows us to see the shapes of the dependencies at low values of $\sigma/F$. 

For a given rms energy dependence, we have to choose the inner boundary condition (specifying the dissipation profile), equations (\ref{d_no_b}) or (\ref{d_b}), and the radial rms profile, equations (\ref{hankel}) (characterized by $2\upi R_{\rm in}/\lambda$), or either (\ref{r_pl}) or (\ref{r_d}) (characterized by $\beta$). For the power-law emissivity, equation (\ref{fpl}), we need to further specify $R_{\rm out}/R_{\rm in}$, $\alpha_{\rm in}$, $\alpha_{\rm out}$, $E_0$ and $E_{\rm c}$. In the cases of incoherent variability, an additional free parameter is the coherence length, $\Delta\!\ln R$ (in the logarithmic case, which we will assume hereafter). 

\subsection{Blackbody and Wien spectra}
\label{blackbody}

\begin{figure}
\centerline{\psfig{file=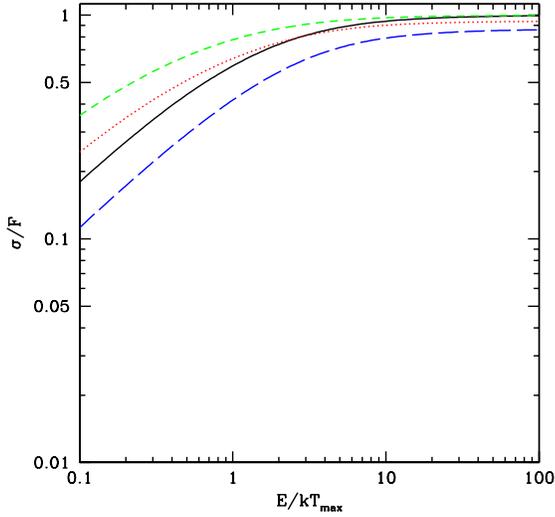,width=7.3cm}} 
\caption{Example rms energy dependencies for the radial rms profile corresponding to disc waves (maintaining coherence). The short dashes and the solid curve correspond to the torqued inner boundary at $2\upi R_{\rm in}/\lambda=0.01$ and 1, respectively, while the dots and long dashes show the corresponding results for the zero-torque boundary. The dependencies for that ratio $\ga 1$ are virtually identical to those for $=1$. 
\label{rms_hankel} }
\end{figure}

We first consider the rms radial profile corresponding to disc waves, equation (\ref{hankel}). For a given dissipation profile, the free parameter is the ratio of the inner disc radius to the wavelength, $2\upi R_{\rm in}/\lambda$, dependence on which disappears for $2\upi R_{\rm in}/\lambda\ga 1$ as $r(R)$ is then a power law with the constant index of $-1/2$ (Section \ref{radial}). Fig.\ \ref{rms_hankel} shows the coherent rms profiles for torqued and zero-torque inner boundaries and $2\upi R_{\rm in}/\lambda=0.01$ and 1.

\begin{figure}
\centerline{\psfig{file=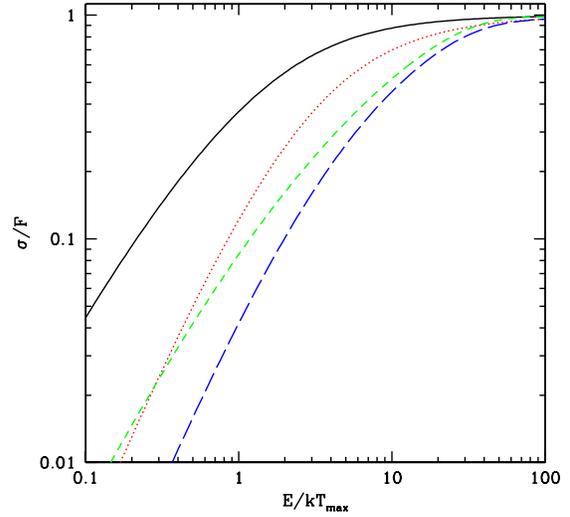,width=7.3cm}} 
\caption{Dependence of the rms on $\beta$. The solid curve and dots show the coherent dependence for $\beta=1$ and 3, respectively. The short and long dashes show the corresponding incoherent dependencies for $\Delta\!\ln R=0.1$. 
\label{rms_beta} }
\end{figure}

\begin{figure}
\centerline{\psfig{file=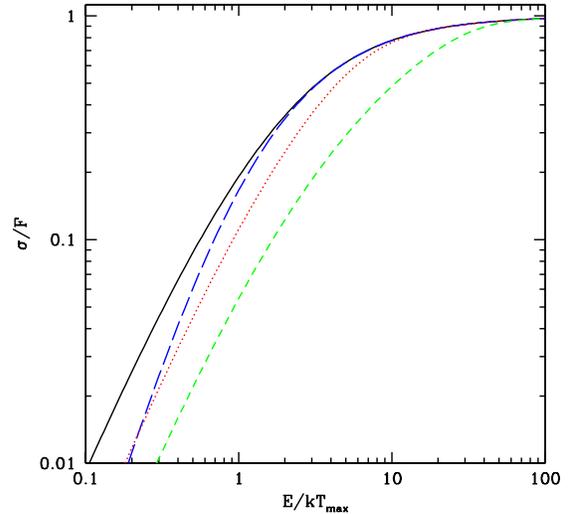,width=7.3cm}} 
\caption{The coherent rms for the blackbody (solid curve) and the Wien spectrum (long dashes) for $\beta=2$, and the dependence of the corresponding incoherent rms on $\Delta\!\ln R$, with the short dashes and dots showing the cases for $\Delta\!\ln R=\ln 1.1$ and $\ln 1.5$, respectively. 
\label{rms_delta} }
\end{figure}

\begin{figure}
\centerline{\psfig{file=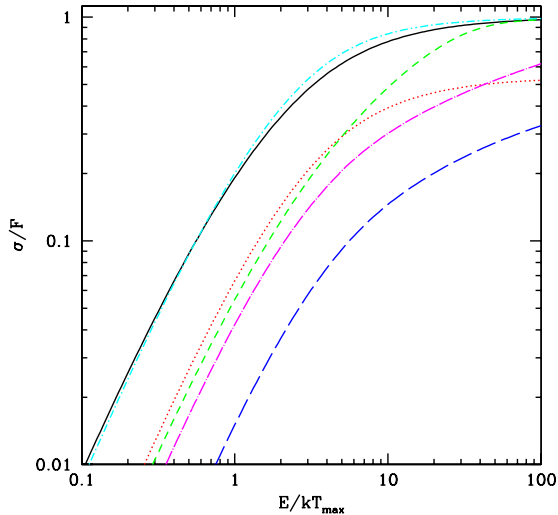,width=7.3cm}} 
\caption{Dependences of rms profiles on the disc inner boundary condition. The solid curve shows the coherent dependence for $\beta=2$ for the dissipation profile with the torqued inner boundary. Then the dots show the corresponding profile for the zero-torque inner boundary and the local rms being a power law in radius, equation (\ref{r_pl}). The dots/short dashes show the analogous profile, but with the local rms following the dissipation rate, equation (\ref{r_d}). The short dashes, long dashes and dots/long dashes show the corresponding incoherent profiles for $\Delta\!\ln R=0.1$. 
\label{rms_t} }
\end{figure}

Note that Figs.\ \ref{rms_b3}--\ref{rms_t} show the values of the rms up to $E=100 kT_{\rm max}$, in order to show the shape of the dependence including their asymptotic behaviour. However, the blackbody emission becomes very weak for $E\ga 15 kT_{\rm max}$, and then other spectral components, e.g., Comptonization (Section \ref{pl}) may dominate. To allow direct comparison of the shape, Figs.\ \ref{rms_hankel}--\ref{rms_t} have the same range of $\sigma/F$.

Fig.\ \ref{rms_beta} shows a comparison between results in the blackbody case with the power-law radial rms dependence and the torqued inner boundary for two values of $\beta$ for both the coherent and incoherent variability. The rms($E$) profiles change much more than in the case of the wave emissivity (Fig.\ \ref{rms_hankel}). The rms decreases with the increasing $\beta$, and the incoherent variability is weaker than the coherent one for a given $\beta$. Note that the rms at $\beta>5/4$ has an asymptotic power-law form with the index of 5/3 at low energies, see Section \ref{analytical}. 

Fig.\ \ref{rms_delta} compares the coherent rms in the case of local Wien emission, equation (\ref{ratio_wien}), with that for the local blackbody for $\beta=2$ and the torqued inner boundary. The blackbody and Wien low-energy power-law indices are 5/3 and $4\beta/3=8/3$, respectively, see see Section \ref{analytical}.

Fig.\ \ref{rms_delta} also shows the dependence of incoherent rms($E$) on $\Delta\!\ln R$. We see the rms increases with the increasing $\Delta\!\ln R$, and it approaches the coherent dependence at a a given $E/kT_{\rm max}$ for a large enough $\Delta\!\ln R$. This is due to the incoherent emission being then dominated by the innermost disc element. 

Then, Fig.\ \ref{rms_t} shows effects of the inner boundary condition. We see that if we assume the zero-torque inner boundary but keep the rms($R$) as a power law, the resuling $\sigma/F$ are reduced by about an order of magnitude with respect to the case with the torqued inner boundary. On the other hand, assuming that the local rms does not directly follow the radius but instead the dissipation rate, equation (\ref{r_d}), leads to the $\sigma/F$ approximately returning to the original values for torqued boundary condtion. Still, the incoherent rms is now lower by a factor of $\sim$2 at high energies than that for the torqued boundary case due to the radial temperature profile being now less peaked. 

\subsection{Thermal Comptonization}
\label{pl}

\begin{figure}
\centerline{\psfig{file=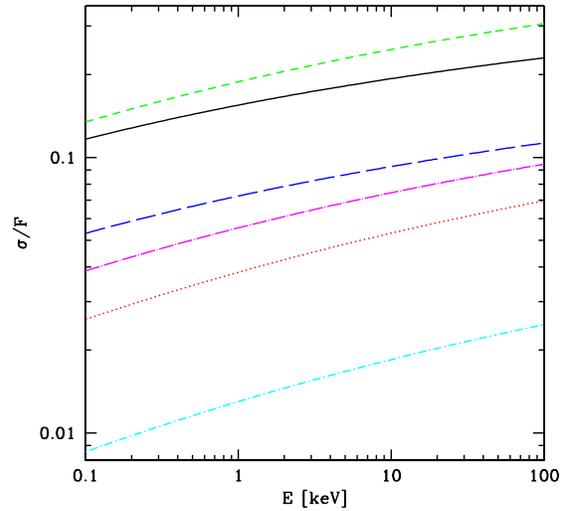,width=7.3cm}} 
\caption{Example rms dependencies for an e-folded power law with the spectral index changing from $\alpha_{\rm out}=0.9$ at $100 R_{\rm in}$ to $\alpha_{\rm in}=0.4$ at $R_{\rm in}$ for $E_0=0.1$ keV, $E_{\rm c}=100$ keV, and $\beta=4$. The solid curve, dots and short dashes correspond to the power-law radial rms profile and the torqued inner boundary, the same profile for the zero-torque boundary, and the the rms profile following the latter dissipation profile, respectively. The long dashes, dots/short dashes and dots/long dashes show the corresponding dependencies, respectively, for incoherent variability with $\Delta\! \ln R=0.1$. 
\label{rms_pl} }
\end{figure}

We then consider the case of the e-folded power law, approximating thermal Comptonization. Fig.\ \ref{rms_pl} shows the rms dependencies for $\beta=4$ and various assumptions regarding the coherence and the profiles of dissipation and the local rms. The overall rms level strongly depends on those assumptions. All the dependencies are increasing, reflecting the assumption that the spectrum hardens with decreasing radius, though much flatter than in the blackbody case. We can obtain the rms increasing with radius for either the spectrum softening or the local variability decreasing with the decreasing radius.

\begin{figure}
\centerline{\psfig{file=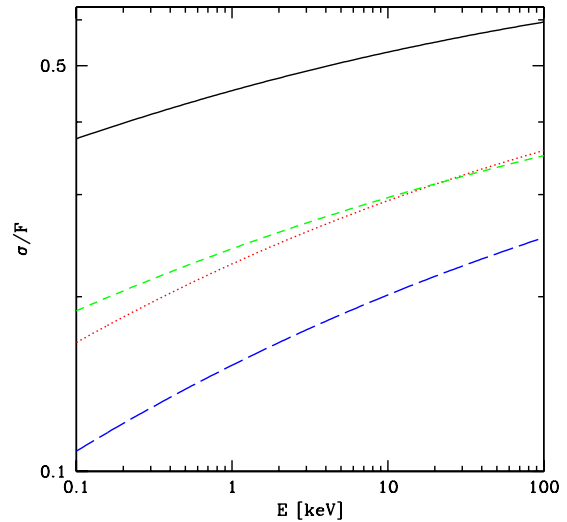,width=7.3cm}} 
\caption{Example rms profiles for an e-folded power law with $\alpha_{\rm in}$, $\alpha_{\rm out}$, $R_{\rm out}/R_{\rm in}$, $E_0$, $E_{\rm c}$ as in Fig.\ \ref{rms_pl}, the dissipation profile with the zero-torque inner boundary and the rms governed by its dissipation profile. The solid curve and dots correspond to coherent variability at $\beta=1$ and 3, respectively. The short and long dashes show the corresponding incoherent variability at $\Delta\! \ln R=\ln 2$. 
\label{rms_pl_beta} }
\end{figure}

Fig.\ \ref{rms_pl_beta} show the effect of varying $\beta$ for both coherent variability and incoherent one with $\Delta\! \ln R=\ln 2$. The increasing $\beta$ results in decreasing rms.

\section{Application to GRS 1915+105}
\label{grs}

\begin{figure}
\centerline{\psfig{file=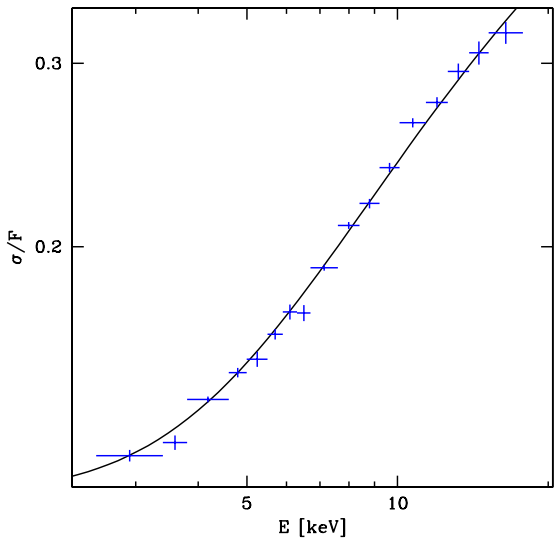,width=7.3cm}} 
\caption{The rms observed from GRS 1915+105 in an ultrasoft, blackbody-dominated, state. The solid curve show the theoretical model with incoherent variability with $\Delta\!\ln R=0.1$ in a flow with a zero-torque inner boundary, the radial power-law rms profile following the dissipation rate at $\beta=4$ and normalized to $r_{\rm max}=1.35$. A weak rms component constant with energy has also been added (see text). 
\label{rms_813} }
\end{figure}

We compare here our theoretical models with local blackbody emission to the rms from an observation of the microquasar GRS 1915+105 performed by the \xte/PCA on 1999 April 21. Its variability properties have been studied in detail in the Zdziarski et al.\ (2005). It belongs to the $\omega$ (Klein-Wolt et al.\ 2002) variability class. It is characterized by a very bright, ultrasoft spectrum, which is well fitted by a disc blackbody Comptonized by a relatively cold, Thomson-thick plasma (Zdziarski et al.\ 2001). Still, we have found that the 3--20 keV spectrum can be approximated within 10 percent by an absorbed disc blackbody alone with the maximum temperature of $kT_{\rm max}\approx 2.3$ keV. Thus, we apply to it the formalism developed above for such spectra. Details of determination of the rms profile are given in Zdziarski et al.\ (2005). Fig.\ \ref{rms_813} shows the 3--18 keV rms, obtained directly from the lightcurve with 1/128 s time bins.

We have performed an extensive search of theoretical models fitting the data. We have found, in particular, that we can rule out models with the rms profile corresponding to the amplitude of a disc wave, equation (\ref{hankel}). The reason for it is its rms($E$) dependence being much too flat (Fig.\ \ref{rms_hankel}) than the observed dependence. The data require, in fact, an rms profile much steeper than $\beta=1/2$ of the wave profile. Our best model, shown in Fig.\ \ref{rms_813}, has $\beta=4$, incoherent variability with $\Delta\!\ln R=0.1$ and the dissipation and rms profiles of equations (\ref{d_b}) and (\ref{r_d}), respectively. The flattening seen at low energies then requires an addition (in quadrature) of a variability pattern with the rms independent of energy (corresponding to a varying normalization of the entire spectrum) at the amplitude of $\sigma_{\rm const}/F= 0.115$. 

We stress we have approximated here the photon spectrum as a disc blackbody, which only roughly approximates the actual spectrum. Taking into account this inaccuracy will modify somewhat the model rms dependencies. Therefore, we
we have not systematically fitted the rms data, e.g., by the least-square method, as the resulting small fit improvements may be less significant than the inaccuracies introduced by our assumption above.

\section{Discussion and conclusions}

We have presented a formalism to calculate rms spectra from accretion flows under the assumptions that each radius is characterized by a given value of the local rms and a given spectral form (Section \ref{disc}). We do not constrain a priori the profiles of the local rms, but it is likely that the flow becomes more turbulent and unstable with decreasing radius, which results in the rms increasing with decreasing $R$ (at least down to some radius). In Section \ref{radial}, we consider three specific radial emissivity profiles. In the case of a local blackbody or Wien spectra, the peak energy increases (at least for most radii) with the decreasing radius, which leads to rms spectra increasing with energy, see Section \ref{blackbody}. This is indeed observed in an example of a blackbody-like emission we consider (Section \ref{grs}). For the case of local Wien emission, we obtain analytical formulae for the integrated disc spectrum, which we find to be a good approximation to the disc blackbody spectrum (Appendix A), and the rms, Section \ref{analytical}. For both blackbody and Wien spectra, we obtain formulae connecting the power-law index, $\beta$, of the radial rms profile with the low-energy index of the energy rms dependence, equal to either $4\beta/3$ or 5/3 (Section \ref{analytical}). Thus, measurements of the indices of rms energy dependence can constrain the variability properties of the emitting accretion disc.

We have also considered the local emission due to thermal Comptonization (Section \ref{radial}). If the seed photons are emitted by an outside cold disk (as is likely to be the case of the hard state of black-hole binaries), the power-law part of the local Comptonization spectrum will harden with the decreasing radius. Such a behaviour is indeed observed in some sources (Revnivtsev et al.\ 1999). Again, integration over radii yields rms spectra increasing with energy (Section \ref{pl}), though with an overall dependence much flatter than in the blackbody case. Such a behaviour is also obtained in models considering in more detail propagation of disturbances towards the central black hole (\.Zycki 2003; P. T. \.Zycki, personal communication).

On the other hand, some black-hole binaries in the hard state show X-ray rms spectra decreasing with energy (e.g., Wardzi\'nski et al.\ 2002). This is likely to be caused by varying shape of the spectrum emitted at a given radius (Section \ref{intro}), which effect is not included in the formalism presented here. 

We have also not considered variability related to the orbital motion at a given radius. This effect may be important observationally in AGNs (Turner et al.\ 2004), where the orbital motion can be well resolved by X-ray observations.

Given the theoretical uncertainty about the level of turbulence in accretion discs as a function of radius and the degree of correlation between emission from different radii, our models rely on a number of assumptions and free parameters, e.g., about the radial profile of the rms, see Section \ref{examples}. Still, data of sufficient quality can strongly constrain the models. For example, blackbody rms energy dependence flattens at a much lower energy in the coherent case than in the incoherent one (Section \ref{blackbody}). 

We stress that rms studies can, and should, be combined with other observational constraints. The components present in time-averaged energy spectra can be determined by spectral fits. Furthermore, the degree of coherence between emission at different energy bands (and, presumably, different radii), can be found by cross-correlation of the corresponding lightcurves. A more detailed diagnostic is the coherence function (Vaughan \& Nowak 1997), the cross-correlation coefficient as a function of Fourier frequency. Also, coherence between spectral components can be tested directly from the rms energy dependence, as the rms$(E)$ from coherently varying components would be simply the weighted average of the rms of the components (Section \ref{comp}). On the other hand, two components varying independently would give a dip in the rms$(E)$ at the energy where their spectra intersect, due to the partial cancellation of their independent variability. Such dips appear, e.g., to be absent in the $\chi$ states of GRS 1915+105 (Zdziarski et al.\ 2005). 

In future work, the various approaches to calculating observed rms spectra can be combined. E.g., more than one component with radial dependencies of rms (Section \ref{comp}) can be considered (Zdziarski et al.\ 2005), or the radial dependence can be combined with time evolution at a given radius. Then, the rms from time evolution of a uniform Comptonization source due to variability of the plasma parameters is considered by Gierli\'nski \& Zdziarski (in preparation). 

\section*{ACKNOWLEDGMENTS}

This research has been supported by KBN grants PBZ-KBN-054/P03/2001, 1P03D01827 and 4T12E04727. I thank Marek Gierli\'nski for valuable discussions, suggestions and calculating the rms spectrum of GRS 1915+105, the referee and Ranjeev Misra for valuable suggestions, and Marat Gilfanov for careful reading of the manuscript.

\appendix
\section{Disc blackbody and disc Wien spectra}
\label{wien}

In the case of the torqued inner boundary, equation (\ref{d_no_b}), and $R_{\rm out}=\infty$, the disc blackbody spectrum (equations [\ref{flux}], [\ref{fbb}]) is,
\begin{equation}
F={8 \upi\over 3} A R_{\rm in}^2 E^3
\cases{\int_1^\infty {t^{5/3} {\rm d}t \over \exp(\epsilon t)-1}, \cr
\epsilon^{-8/3} \Gamma \! \left(8\over 3\right)
\zeta\! \left(8\over 3\right), &$\epsilon\ll 1$;\cr
\epsilon^{-1} \exp(-\epsilon), &$\epsilon\gg 1$,\cr}
\label{diskbb}
\end{equation}
where $\epsilon \equiv E/kT_{\rm max}$, $\zeta(8/3)\simeq 1.28419$, and $\Gamma (8/3)\simeq 1.50458$. 

On the other hand, the blackbody spectrum in the Wien limit becomes,
\begin{equation}
F(R)=A E^3 \exp[-E/kT(R)].
\label{fwien}
\end{equation}
This is actually the equilibrium photon distribution reached in a purely scattering medium (without stimulated scattering) with infinite optical depth, and thus it may be a better approximation to emission of effectively-thin (but optically thick to scattering) accretion discs (Shakura \& Sunyaev 1973) than the blackbody distribution. Wien-emitting discs have been considered, e.g., by Misra, Chitnis \& Melia (1998) and Misra (2000).

\begin{figure}
\centerline{\psfig{file=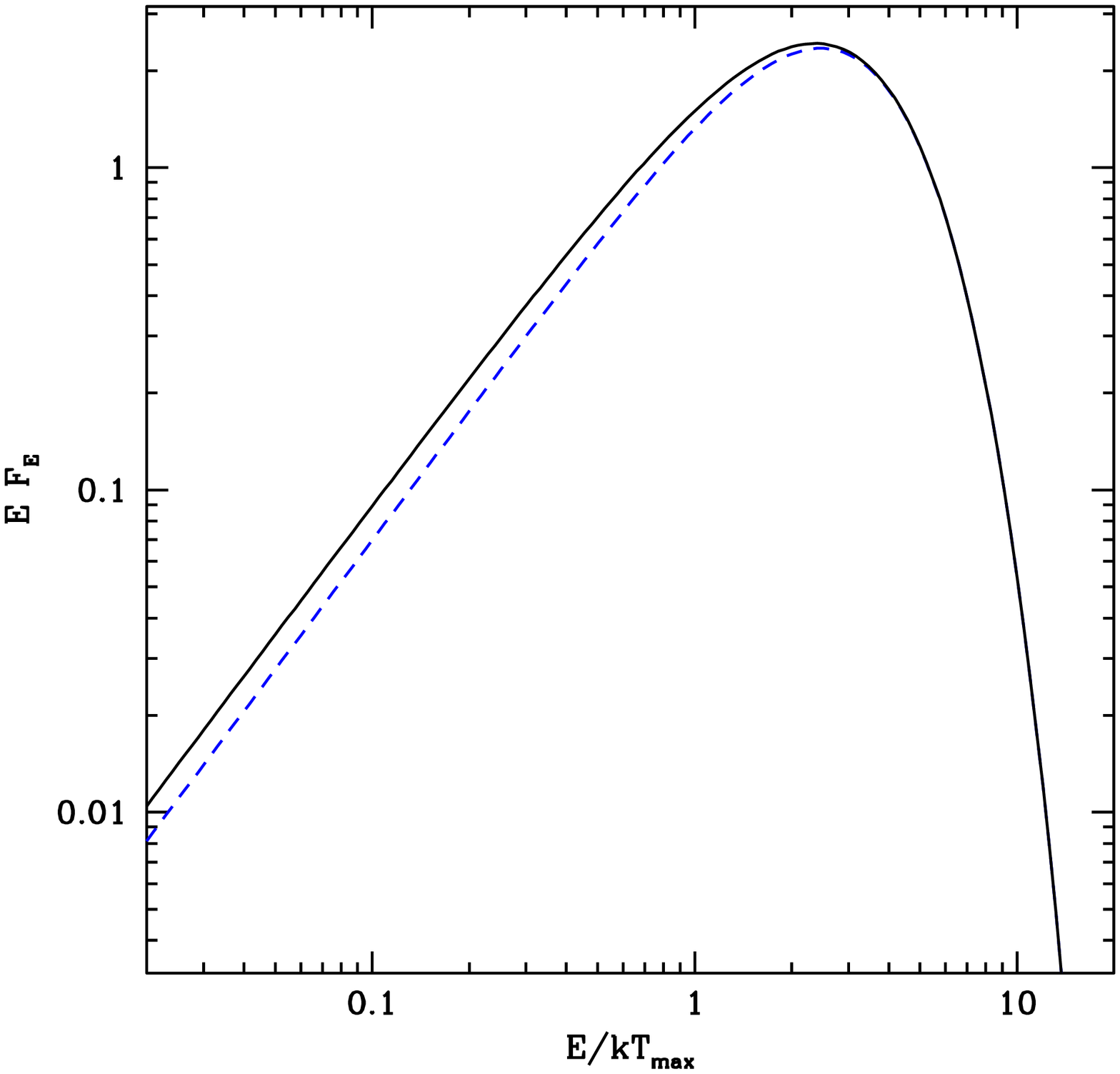,width=7.3cm}} 
\caption{The disc blackbody spectrum (solid curve) and the disc Wien spectrum (dashes) for the torqued inner boundary (in the $EF_E$ representation).
\label{dbb_wien} }
\end{figure}

With the same assumptions as for equation (\ref{diskbb}), the disc Wien spectrum can be expressed in a closed form, 
\begin{equation}
F= {8 \upi\over 3} A R_{\rm in}^2 E^3 \epsilon^{-8/3} \Gamma \left({8\over 3}, \epsilon \right),
\label{disc_wien}
\end{equation}
Notably, this spectrum is very close to the blackbody spectrum, see Fig.\ \ref{dbb_wien}. Its limiting forms are the same as those in equation (\ref{diskbb}) except for no $\zeta(8/3)$ factor at $\epsilon\ll 1$.

\label{lastpage}
\end{document}